\documentclass[aps,prd,twocolumn,nofootinbib]{revtex4-2}
\usepackage[T1]{fontenc}
\usepackage[utf8]{inputenc}
\usepackage[english]{babel}
\usepackage[hidelinks]{hyperref}
\usepackage{microtype,amsfonts,amssymb,amstext,amsmath,mathtools,cancel,physics}
\newcommand{\rvline}{\hspace*{-\arraycolsep}\vline\hspace*{-\arraycolsep}}
\DeclareMathOperator{\sign}{sgn\!}
\DeclareMathOperator{\diag}{diag}

\begin{document}
\title{Pre-geometric Einstein--Cartan Field Equations and Emergent Cosmology}

\author{Giuseppe Meluccio\,}
	\email{giuseppe.meluccio-ssm@unina.it}
	\affiliation{Scuola Superiore Meridionale, Largo San Marcellino, 10, 80138, Napoli, Italy}
	\affiliation{INFN Sezione di Napoli, Complesso Universitario di Monte Sant'Angelo, Edificio 6, Via Cintia, 80126, Napoli, Italy}

\date{\today}

\begin{abstract}
	The field equations of pre-geometric theories of gravity are derived and analysed, both without and with matter. After the spontaneous symmetry breaking that reduces the gauge symmetry of these theories à la Yang--Mills, a metric structure for spacetime emerges and the field equations recover both the Einstein and the Cartan field equations for gravity. A first exact solution of the pre-geometric field equations is also presented. This solution can be considered as a pre-geometric de Sitter universe and provides a possible resolution for the problem of the Big Bang singularity.
\end{abstract}

\maketitle

\section{Introduction}\label{sec:1}
A deeper understanding of gravity is linked to multiple topics of present-day fundamental Physics, especially Quantum Gravity and Cosmology. A widespread idea in this regard is that of the emergence of spacetime from some more primitive structure, which could possibly help to bridge the gap between the geometric description of General Relativity and the probabilistic one of the quantum theory \cite{thiemann:modern,barrett:lorentzian,ashtekar:review,seiberg:emergent,yang:emergent,van-raamsdonk:entanglement,maitiniyazi:irreversible,maitiniyazi:generation}. Pre-geometry is then the idea of constructing a more fundamental model of spacetime which can account for the classical metric structure of Einstein gravity as an emergent phenomenon \cite{akama:pregeometry,akama:topological,wetterich:spinors,wetterich:cosmology,wetterich:pregeometry,akama:pregeometric,floreanini:topological}. A key challenge in this context is the implementation of the gauge principle into any gravitational theory \cite{utiyama:invariant,cai:f(T),hehl:spin,gronwald:gauge}. In particular, Gauge gravitation theory tries to tackle this issue by extending the Yang--Mills formalism to a gauge theory with a non-compact gauge group and thus to gravity \cite{ivanenko:gauge,sardanashvily:gauge,hehl:metric,blagojević:gravitation,blagojević:gauge,leclerc:higgs,randono:gauge,ponomarev:gauge,cabral:gauge,celada:bf,de-pietri:plebanski,buffenoir:hamiltonian,durka:hamiltonian}. An important insight for achieving this goal, inspired by the Higgs mechanism of particle Physics, is that the phenomenon of spontaneous symmetry breaking (SSB) can play a role not only for the quantum interactions, but also for the gravitational one \cite{adler:einstein,alexander:higgs,percacci:higgs,volovik:phase}.

The pre-geometric theories of gravity studied in this article attempt to reconcile all of the above features: starting from a generally covariant theory à la Yang--Mills for a gauge field $A$ formulated in a four-dimensional spacetime endowed with no metric structure, the emergence of gravity and geometry is then dynamically induced by the SSB of the non-compact gauge group of the theory \cite{macdowell:unified,chamseddine:supergravity,stelle:de-sitter,wilczek:gauge,westman:cartan,westman:dynamical,westman:introduction,pagels:gravitational,mccarthy:surface,mielke:spontaneously,tresguerres:dynamically,wise:cartan,lisi:unification,smolin:holographic,gallagher:pregeometric,addazi:pre-geometry,addazi:hamiltonian}. Such gauge group can be taken to be either the de Sitter group $SO(1,4)$ or the anti-de Sitter group $SO(2,3)$. The phase transition from the unbroken to the spontaneously broken phase for spacetime, i.e.\ from a pre-geometric to a metric universe, is realised at a critical energy (near the Planck scale) by the dynamics of a Higgs-like field $\phi$. As a result of the SSB, the original gauge symmetry is reduced to that of the Lorentz group $SO(1,3)$. Despite being unrelated to gravitation in principle, this pre-geometric framework can thus provide a full characterisation of gravity as an emergent phenomenon: its dynamics in the form of the Einstein--Cartan theory, its mass parameters (the Planck scale and the cosmological constant) and its physical principles (background independence and the equivalence principle) are all recovered after the SSB of the vacuum state of the initial gauge theory. Under very general assumptions, one can prove that only two such theories exist \cite{addazi:pre-geometry}, whose Lagrangian densities were first written down by MacDowell and Mansouri \cite{macdowell:unified} and Wilczek \cite{wilczek:gauge} respectively. For both theories, the gravitational Higgs mechanism was illustrated in Ref.\ \cite{addazi:pre-geometry}, while the Hamiltonian analysis -- which is of paramount importance for quantisation -- was carried out in Ref.\ \cite{addazi:hamiltonian}. The main result of the Hamiltonian analysis is that the number of degrees of freedom is three both for the Wilczek and the MacDowell--Mansouri theories. This should be contrasted with the Einstein--Cartan theory of gravity, whose number of degrees of freedom is two, corresponding to a massless graviton. In the case of pre-geometric theories of gravity, in fact, in addition to a massless graviton there is also an extra scalar field, corresponding to the physical quanta of the Higgs-like field $\phi$.

So far, pre-geometric theories have mostly been analysed in their Lagrangian formulation in the literature, with little or no attention paid to their equations of motion or solutions. Given the interesting properties of these theories, their equations of motion hold promise for results that could illuminate at least some of the theoretical problems of General Relativity. Therefore, in this work we examine the field equations of pre-geometric theories, both without and with matter. The latter case requires supposing that matter couples to the pre-geometry of spacetime in the unbroken phase just as it does to its geometry in the spontaneously broken phase. In doing so, emphasis is placed on the way the field equations of the Einstein--Cartan theory are recovered after the SSB, but also on the features and the consequences of the pre-geometric field equations in the unbroken phase. In particular, the latter regime points at a possible unification of the dynamics involving the metric and the affine structures of spacetime on one side and the energy-momentum and spin properties of matter on the other. In addition to ascertaining the consistency of the low-energy limit, a first exact solution of the pre-geometric field equations is presented. In the ultra-high-energy regime where the fundamental gauge symmetry of spacetime is restored, this solution can be thought of as a pre-geometric de Sitter spacetime, which also provides a novel solution for the problem of the Big Bang singularity.

The internal space metric $\eta$ used to define the gauge groups $SO(1,4)$ or $SO(2,3)$ is a generalisation of the Minkowski metric and its signature is respectively $(-,+,+,+,\pm)$. Whenever a double sign is encountered in this work, the convention is to refer the first sign to the case of $SO(1,4)$ and the second sign to the case of $SO(2,3)$, unless otherwise specified. After the SSB, both gauge groups reduce to $SO(1,3)$ and the internal space metric $\eta$ becomes the Minkowski metric with signature $(-,+,+,+)$. Tangent space (or internal) indices are represented by uppercase Latin letters $A,B,C$ etc.\ in the unbroken phase or lowercase Latin letters $a,b,c$ etc.\ in the spontaneously broken phase; these indices run from $0$ to $4$ or from $0$ to $3$ respectively. Spacetime (or external) indices are represented by Greek letters $\lambda,\mu,\nu$ etc.\ and run from $0$ to $3$, with lowercase Latin letters $i,j,k$ etc.\ denoting spatial indices and running from $1$ to $3$ instead. Natural units are used throughout the article.

The paper is organised as follows. In Sec.\ \ref{sec:2} we summarise the essentials of the formalism of the Einstein--Cartan theory. The equations of motion of pre-geometric theories in the absence of matter are discussed in Sec.\ \ref{sec:3} and then solved in a cosmological setting in Sec.\ \ref{sec:4}. The results of these two sections are then generalised respectively in Secs.\ \ref{sec:5} and \ref{sec:6} with the introduction of the matter coupling, leading to the general form of the pre-geometric field equations and their cosmological application. Lastly, Sec.\ \ref{sec:7} provides some concluding remarks as well as perspectives for further developments.

\section{The formalism of the Einstein--Cartan theory}\label{sec:2}
The Einstein--Cartan theory describes the gravitational interaction as the effect of both the curvature and the torsion of spacetime. In the metric formulation, the independent variables of the spacetime geometry are the metric $g$ and the torsion $T$ (or the affine connection $\Gamma$). Analogously, in the tetrad formulation, which is also referred to as the Sciama--Kibble theory \cite{kibble:lorentz,sciama:physical,capozziello:torsion,popławski:classical,di-stefano:canonical,castellani:tetrad}, the independent variables can be taken to be the tetrads $e$ and the spin connection $\omega$. The fundamental relation linking the metric and the tetrads is
\begin{equation}\label{eq:soldered}
	g_{\mu\nu}=\eta_{ab}e_\mu^ae_\nu^b,
\end{equation}
where $\eta$ is the Minkowski metric. Just as the metric tensor allows to raise and lower spacetime indices, so the tetrad fields allow to convert spacetime indices into tangent space ones or vice versa, thanks to the soldered property of spacetime expressed by Eq.\ \eqref{eq:soldered}. In terms of the tetrads and the spin connection, curvature and torsion are defined respectively as
\begin{gather}
	R_{\mu\nu}^{ab}=2\partial_{[\mu}\omega_{\nu]}^{ab}+2\omega_{c[\mu}^a\omega_{\nu]}^{cb},\\
	T_{\mu\nu}^a=2\partial_{[\nu}e_{\mu]}^a+2\omega_{b[\nu}^ae_{\mu]}^b=2\mathcal{D}_{[\nu}e_{\mu]}^a,
\end{gather}
where $\omega_{b\mu}^a\equiv\omega_{\phantom{b}b\mu}^a=\eta_{bc}\omega_\mu^{ac}$ and $\mathcal{D}$ denotes the covariant derivative with respect to $\omega$.

The total action of the theory is
\begin{equation}\label{eq:S-geometry}
	S=S_\textup{EC}[e,\partial e,\omega,\partial\omega]+S_m[e,\partial e,\omega,\psi_n,\partial\psi_n],
\end{equation}
where the action for $n$ matter fields $\psi_n$ with total Lagrangian density $\mathcal{L}_m$ is defined as
\begin{equation}
	S_m=\int e\mathcal{L}_m\,d^4x
\end{equation}
and the Einstein--Cartan action for gravity is defined as the Einstein--Hilbert action in the tetrad formalism with the addition of the cosmological constant term:\footnote{We have included $\partial e$ in the definition of $S_\textup{EC}$ for more generality: even though no derivatives of the tetrads appear in the expressions of Eq.\ \eqref{eq:EC}, in principle they can appear if at least some of the terms containing $\partial\omega$ in $R$ are integrated by parts.}
\begin{equation}\label{eq:EC}
	\begin{split}
		S_\textup{EC}&=\frac{M_\textup{P}^2}{2}\int e(e_a^\mu e_b^\nu R_{\mu\nu}^{ab}-2\Lambda)\,d^4x\\
		&=-\frac{M_\textup{P}^2}{8}\epsilon_{abcd}\epsilon^{\mu\nu\rho\sigma}\int\biggl(R_{\mu\nu}^{ab}-\frac{\Lambda}{3}e_\mu^ae_\nu^b\biggr)e_\rho^ce_\sigma^d\,d^4x,
	\end{split}
\end{equation}
where $M_\textup{P}$ is the reduced Planck mass, $\Lambda$ is the cosmological constant and $e\equiv\det(e_\mu^a)$. The equivalence between the two expressions shown in Eq.\ \eqref{eq:EC} is valid whenever the tetrads are invertible;\footnote{For $e\ne0$, the identity $\epsilon_{abcd}\epsilon^{\mu\nu\rho\sigma}R_{\mu\nu}^{ab}e_\rho^ce_\sigma^d=-4ee_a^\mu e_b^\nu R_{\mu\nu}^{ab}$ holds true \cite{addazi:pre-geometry}.} the advantage of working with the second rather than the first expression lies in never having to resort to the inverse of the tetrad fields. Note that all terms of $S$ containing $\partial\omega$ belong to the gravitational part $S_\textup{EC}$ of the action and not to the matter part $S_m$.

The variational principle $\delta S=0$ must be applied to both independent variables of the action \eqref{eq:S-geometry}. The variation of $S$ with respect to the tetrads yields the Einstein field equations,
\begin{equation}\label{eq:einstein-eq}
	G_a^\mu=M_\textup{P}^{-2}\tau_a^\mu,
\end{equation}
where the tensor on the l.h.s.\ is the Einstein tensor
\begin{equation}\label{eq:einstein-t-i}
	G_a^\mu\equiv\frac{1}{4e}\epsilon_{abcd}\epsilon^{\mu\lambda\rho\sigma}\biggl(R_{\lambda\rho}^{bc}-\frac{2\Lambda}{3}e_\lambda^be_\rho^c\biggr)e_\sigma^d
\end{equation}
and that on the r.h.s.\ is the canonical energy-momentum tensor
\begin{equation}
	\tau_a^\mu\equiv\frac{1}{e}\fdv{(e\mathcal{L}_m)}{e_\mu^a}.
\end{equation}
Multiplication of Eq.\ \eqref{eq:einstein-eq} by $e_\nu^a$ recasts the Einstein field equations in the form with spacetime indices only which is familiar from the metric formulation of the theory,
\begin{equation}
	G_\nu^\mu=M_\textup{P}^{-2}\tau_\nu^\mu,
\end{equation}
where the definition of the Einstein tensor is inclusive of the cosmological constant term for convenience:\footnote{The expression \eqref{eq:einstein-t-e} of the Einstein tensor can be derived from that of Eq.\ \eqref{eq:einstein-t-i} by means of the identities $\epsilon_{abcd}\epsilon^{\mu\lambda\rho\sigma}e_\nu^ae_\lambda^be_\rho^ce_\sigma^d=-6e\delta_\nu^\mu$ and $\epsilon_{abcd}\epsilon^{\mu\lambda\rho\sigma}e_\nu^aR_{\lambda\rho}^{bc}e_\sigma^d=4eR_\nu^\mu-2e\delta_\nu^\mu R$. See Ref.\ \cite{addazi:pre-geometry} for details on how to derive these identities.}
\begin{equation}\label{eq:einstein-t-e}
	G_\nu^\mu\equiv R_\nu^\mu-\frac{1}{2}\delta_\nu^\mu R+\Lambda\delta_\nu^\mu.
\end{equation}

The variation of $S$ with respect to the spin connection yields the Cartan field equations,\footnote{Two useful identities to compute the variation of $S_\textup{EC}$ with respect to $\omega$ are $\delta R_{\mu\nu}^{ab}=2\mathcal{D}_{[\mu}\delta\omega_{\nu]}^{ab}$ and $\mathcal{D}_\mu e=ee_a^\nu\mathcal{D}_\mu e_\nu^a$.}
\begin{equation}\label{eq:cartan-eq}
	\tilde{T}_{ab}^\mu=M_\textup{P}^{-2}\sigma_{ab}^\mu,
\end{equation}
where the tensor on the l.h.s.\ is the modified torsion tensor
\begin{equation}\label{eq:modified-torsion-t-i}
	\begin{split}
		\tilde{T}_{ab}^\mu&\equiv\frac{2}{e}\mathcal{D}_\lambda(ee_{[a}^\lambda e_{b]}^\mu)\\
		&=\frac{1}{2e}\epsilon_{abcd}\epsilon^{\mu\lambda\xi\sigma}\mathcal{D}_\lambda(e_\xi^ce_\sigma^d)\\
		&=\frac{1}{2e}\epsilon^{\mu\lambda\xi\sigma}(2\epsilon_{cde[a}\omega_{b]\lambda}^e+\epsilon_{abcd}\partial_\lambda)(e_\xi^ce_\sigma^d)
	\end{split}
\end{equation}
and that on the r.h.s.\ is the canonical spin (current) tensor
\begin{equation}
	\sigma_{ab}^\mu\equiv\frac{2}{e}\fdv{(e\mathcal{L}_m)}{\omega_\mu^{ab}}=\frac{2}{e}\pdv{(e\mathcal{L}_m)}{\omega_\mu^{ab}}.
\end{equation}
Multiplication of Eq.\ \eqref{eq:cartan-eq} by $e_\nu^ae_\rho^b$ recasts the Cartan field equations in the form with spacetime indices only,
\begin{equation}\label{eq:cartan-field-eq}
	\tilde{T}_{\nu\rho}^\mu=M_\textup{P}^{-2}\sigma_{\nu\rho}^\mu,
\end{equation}
where the modified torsion tensor is defined in terms of the torsion tensor as\footnote{The expression \eqref{eq:modified-torsion-t-e} of the modified torsion tensor can be derived from that of Eq.\ \eqref{eq:modified-torsion-t-i} by means of the identity $\epsilon_{abcd}\epsilon^{\mu\lambda\xi\sigma}e_\nu^ae_\rho^bT_{\lambda\xi}^ce_\sigma^d=-2e\tilde{T}_{\nu\rho}^\mu$. See Ref.\ \cite{addazi:pre-geometry} for details on how to derive this identity.}
\begin{equation}\label{eq:modified-torsion-t-e}
	\tilde{T}_{\nu\rho}^\mu\equiv T_{\nu\rho}^\mu+2\delta_{[\nu}^\mu T_{\rho]\sigma}^\sigma.
\end{equation}
The Cartan field equations are algebraic equations for the components of the torsion tensor. This implies that the torsion of spacetime, unlike its metric, does not propagate in this theory.\footnote{Despite torsion being non-dynamical in the Einstein--Cartan theory, it can induce four-fermion interactions that have important consequences for cosmological models \cite{alexander:spectrum,alexander:curvaton,addazi:dark,addazi:ekpyrotic,addazi:emergent}.}

\section{Pre-geometric dynamics in vacuum}\label{sec:3}
In this section we discuss the equations of motion for two pre-geometric theories of gravity in the absence of matter in both the unbroken and the spontaneously broken phases. The Lagrangian density for the Wilczek theory is
\begin{equation}\label{eq:L-W}
	\mathcal{L}_\textup{W}=k_\textup{W}\epsilon_{ABCDE}\epsilon^{\mu\nu\rho\sigma}F_{\mu\nu}^{AB}\nabla_\rho\phi^C\nabla_\sigma\phi^D\phi^E,
\end{equation}
where $k_\textup{W}$ is a coupling constant with $[k_\textup{W}]=[\phi]^{-3}$, $F$ is the field strength of the gauge potential $A$,
\begin{equation}\label{eq:F}
	F_{\mu\nu}^{AB}=2\partial_{[\mu}A_{\nu]}^{AB}+2A_{C[\mu}^AA_{\nu]}^{CB}
\end{equation}
with $A_{B\mu}^A\equiv A_{\phantom{B}B\mu}^A=\eta_{BC}A_\mu^{AC}$, and $\nabla$ denotes the covariant derivative with respect to $A$, which acts on the Higgs-like field $\phi$ as
\begin{equation}
	\nabla_\mu\phi^A=\partial_\mu\phi^A+A_{B\mu}^A\phi^B.
\end{equation}
The gauge potential $A$ is antisymmetric in its internal indices and the gauge field strength $F$ is antisymmetric in both its internal and external indices. The Euler--Lagrange equations for $\mathcal{L}_\textup{W}$ with respect to the gauge potential $A$ and the Higgs-like field $\phi$ are respectively
\begin{equation}\label{eq:EoM-A-W}
	\begin{split}
		\epsilon^{\mu\nu\rho\sigma}[&\epsilon_{ABCD[E}\eta_{F]G}(F_{\mu\nu}^{AB}\phi^G-2A_\mu^{AG}\nabla_\nu\phi^B)\nabla_\rho\phi^C\phi^D\\
		&+\epsilon_{ABCEF}\partial_\rho(\nabla_\mu\phi^A\nabla_\nu\phi^B\phi^C)]=0
	\end{split}
\end{equation}
and
\begin{equation}\label{eq:EoM-phi-W}
	\begin{split}
		\epsilon^{\mu\nu\rho\sigma}[&(2\epsilon_{ABCDF}A_{E\sigma}^D\phi^F+\epsilon_{ABCDE}\nabla_\sigma\phi^D)F_{\mu\nu}^{AB}\nabla_\rho\phi^C\\
		&+2\epsilon_{ABCDE}\partial_\sigma(F_{\mu\nu}^{AB}\nabla_\rho\phi^C\phi^D)]=0.
	\end{split}
\end{equation}

The SSB of the vacuum state in the Wilczek theory is realised when the Higgs-like field acquires a nonzero vacuum expectation value along one of the five directions of the internal space;\footnote{Different dynamical mechanisms that can cause the Higgs-like field to acquire a symmetry-breaking vacuum expectation value are discussed in Refs.\ \cite{wilczek:gauge,addazi:pre-geometry,addazi:hamiltonian,addazi:topological}.} choosing such direction as the one with coordinate $A=4$ leads to the following results (see Ref.\ \cite{addazi:pre-geometry} for the details):
\begin{equation}\label{eq:SSB}
	\begin{split}
		\phi^A&\xrightarrow{SSB}v\delta_4^A,\qquad A_\mu^{ab}\xrightarrow{SSB}\omega_\mu^{ab},\qquad A_\mu^{a4}\xrightarrow{SSB}me_\mu^a,\\
		F_{\mu\nu}^{AB}&\xrightarrow{SSB}R_{\mu\nu}^{ab}\mp2m^2e^a_{[\mu}e^b_{\nu]},\qquad\nabla_\lambda\phi^A\xrightarrow{SSB}\pm vme^a_\lambda,
	\end{split}
\end{equation}
where $v=\langle\phi^4\rangle$ and $m$ is a mass parameter introduced in order to define the emergent tetrad fields as dimensionless. Plugging the definitions \eqref{eq:SSB} into $\mathcal{L}_\textup{W}$ yields
\begin{equation}
	\mathcal{L}_\textup{W}\xrightarrow{SSB}k_\textup{W}v^3m^2\epsilon_{abcd}\epsilon^{\mu\nu\rho\sigma}e_\mu^ae_\nu^bR_{\rho\sigma}^{cd}\pm48k_\textup{W}v^3m^4e
\end{equation}
with $\epsilon_{abcd}\equiv\epsilon_{ABCD4}$. By identifying the reduced Planck mass and the cosmological constant respectively as
\begin{equation}\label{eq:emergent-constants-W}
	M_\textup{P}^2\equiv-8k_\textup{W}v^3m^2,\qquad\Lambda\equiv\pm6m^2,
\end{equation}
the Einstein--Cartan action can thus be seen to emerge from the SSB of the Wilczek action:
\begin{equation}\label{eq:SSB-W}
	S_\textup{W}=\int\mathcal{L}_\textup{W}\,d^4x\xrightarrow{SSB}S_\textup{EC}.
\end{equation}

The Lagrangian density for the MacDowell--Mansouri theory is
\begin{equation}\label{eq:L-MM}
	\mathcal{L}_\textup{MM}=k_\textup{MM}\epsilon_{ABCDE}\epsilon^{\mu\nu\rho\sigma}F_{\mu\nu}^{AB}F_{\rho\sigma}^{CD}\phi^E,
\end{equation}
where $k_\textup{MM}$ is a coupling constant with $[k_\textup{MM}]=[\phi]^{-1}$ and $F$ was defined in Eq.\ \eqref{eq:F}. The Euler--Lagrange equations for $\mathcal{L}_\textup{MM}$ with respect to the gauge potential $A$ and the Higgs-like field $\phi$ are respectively
\begin{equation}\label{eq:EoM-A-MM}
	\epsilon^{\mu\nu\rho\sigma}[2\epsilon_{ABCD[E}A_{F]\mu}^AF_{\nu\rho}^{BC}\phi^D-\epsilon_{ABCEF}\partial_\rho(F_{\mu\nu}^{AB}\phi^C)]=0
\end{equation}
and
\begin{equation}\label{eq:EoM-phi-MM}
	\epsilon_{ABCDE}\epsilon^{\mu\nu\rho\sigma}F_{\mu\nu}^{AB}F_{\rho\sigma}^{CD}=0.
\end{equation}

As for the SSB of the vacuum state of the MacDowell--Mansouri theory, the application to $\mathcal{L}_\textup{MM}$ of the same conventions introduced in Eq.\ \eqref{eq:SSB} yields
\begin{equation}
	\begin{split}
		\mathcal{L}_\textup{MM}\xrightarrow{SSB}&\mp4k_\textup{MM}vm^2\epsilon_{abcd}\epsilon^{\mu\nu\rho\sigma}e_\mu^ae_\nu^bR_{\rho\sigma}^{cd}\\
		&-96k_\textup{MM}vm^4e-4k_\textup{MM}veG,
	\end{split}
\end{equation}
where the Euler characteristic of the spacetime manifold is given by
\begin{equation}
	\begin{split}
		G&\equiv R^2-4R_{\mu\nu}R^{\mu\nu}+R_{\mu\nu\rho\sigma}R^{\mu\nu\rho\sigma}\\
		&=(e_a^\mu e_b^\nu e_c^\rho e_d^\sigma-4e_a^\mu e_d^\nu e_c^\rho e_b^\sigma+e_c^\mu e_d^\nu e_a^\rho e_b^\sigma)R_{\mu\nu}^{ab}R_{\rho\sigma}^{cd}\\
		&=-\frac{1}{4e}\epsilon_{abcd}\epsilon^{\mu\nu\rho\sigma}R_{\mu\nu}^{ab}R_{\rho\sigma}^{cd}.
	\end{split}
\end{equation}
By identifying the reduced Planck mass and the cosmological constant this time respectively as
\begin{equation}\label{eq:emergent-constants-MM}
	M_\textup{P}^2\equiv\pm32k_\textup{MM}vm^2,\qquad\Lambda\equiv\pm3m^2,
\end{equation}
the Einstein--Cartan action, supplemented with the Gauss--Bonnet term, can thus be seen to emerge from the SSB of the MacDowell--Mansouri action:
\begin{equation}\label{eq:SSB-MM}
	S_\textup{MM}=\int\mathcal{L}_\textup{MM}\,d^4x\xrightarrow{SSB}S_\textup{EC}+S_\textup{GB},
\end{equation}
with the Gauss--Bonnet term defined as
\begin{equation}
	S_\textup{GB}\equiv\lambda\int eG\,d^4x,\qquad\lambda\equiv-4k_\textup{MM}v,
\end{equation}
where $\lambda$ is an emergent dimensionless coupling constant. For later convenience, we observe that the following relation holds true for the three emergent coupling constants of the MacDowell--Mansouri theory:
\begin{equation}\label{eq:relation-constants-MM}
	\frac{4\lambda}{M_\textup{P}^2}=-\frac{3}{2\Lambda}.
\end{equation}

Since $S_\textup{GB}$ is a topological invariant in less than five spacetime dimensions, it does not contribute to the Einstein field equations, hence the addition of $S_\textup{GB}$ to $S_\textup{EC}$ is irrelevant for the application of the variational principle with respect to the tetrads (or the metric). Conversely, when it comes to computing the variation with respect to the spin connection, the addition of $S_\textup{GB}$ to $S_\textup{EC}$ modifies the l.h.s.\ of the Cartan field equations as follows:
\begin{equation}
	\bar{T}_{ab}^\mu=M_\textup{P}^{-2}\sigma_{ab}^\mu,
\end{equation}
where\footnote{A useful identity to compute the variation of $S_\textup{GB}$ with respect to $\omega$ is $\delta G=2(R_{ab}^{\mu\nu}-4e_a^\mu R_b^\nu+e_a^\mu e_b^\nu R)\delta R_{\mu\nu}^{ab}$.}
\begin{equation}
	\begin{split}
		\bar{T}_{ab}^\mu&\equiv\frac{2}{e}\mathcal{D}_\lambda[ee_{[a}^\lambda e_{b]}^\mu+4\lambda M_\textup{P}^{-2}e(R_{ab}^{\lambda\mu}-4e_{[a}^{[\lambda}R_{b]}^{\mu]}+e_{[a}^\lambda e_{b]}^\mu R)]\\
		&=\frac{1}{2e}\epsilon_{abcd}\epsilon^{\mu\lambda\xi\sigma}\mathcal{D}_\lambda(e_\xi^ce_\sigma^d+4\lambda M_\textup{P}^{-2}R_{\xi\sigma}^{cd})\\
		&=\frac{1}{2e}\epsilon^{\mu\lambda\xi\sigma}(2\epsilon_{cde[a}\omega_{b]\lambda}^e+\epsilon_{abcd}\partial_\lambda)(e_\xi^ce_\sigma^d+4\lambda M_\textup{P}^{-2}R_{\xi\sigma}^{cd}).
	\end{split}
\end{equation}
The difference between the tensors $\tilde{T}$ and $\bar{T}$ is thus linear with respect to $R$. For what concerns the propagation of torsion in this theory, see for example Refs.\ \cite{ozer:modified,niu:torsion}.

After the SSB, the components of the gauge field $A$ describe the emergent tetrads and spin connection according to the definitions of Eq.\ \eqref{eq:SSB}. This fact, together with the results \eqref{eq:SSB-W} and \eqref{eq:SSB-MM}, leads to the following conclusion: for each pre-geometric theory under consideration, the SSB of the Euler--Lagrange equations for $A$ should reproduce both the vacuum Einstein field equations and the vacuum Cartan field equations. The validity of this claim can be proven in a direct way by employing the results \eqref{eq:SSB}, \eqref{eq:emergent-constants-W}, \eqref{eq:emergent-constants-MM} and \eqref{eq:relation-constants-MM}. The SSB of Eqs.\ \eqref{eq:EoM-A-W} and \eqref{eq:EoM-A-MM} is given respectively by
\begin{equation}\label{eq:EoM-A-W-SSB}
	\begin{split}
		\epsilon^{\mu\nu\rho\sigma}\biggl[&\pm\epsilon_{abc[E}\eta_{F]4}\biggl(R_{\mu\nu}^{ab}-\frac{\Lambda}{3}e_\mu^ae_\nu^b\biggr)e_\rho^c\\
		&-2m\epsilon_{abc[E}A_{F]\mu}^ae_\nu^be_\rho^c-m\epsilon_{abEF}\partial_\rho(e_\mu^ae_\nu^b)\biggr]=0
	\end{split}
\end{equation}
and
\begin{equation}\label{eq:EoM-A-MM-SSB}
	\epsilon^{\mu\nu\rho\sigma}(2\epsilon_{abc[E}A_{F]\rho}^c+\epsilon_{abEF}\partial_\rho)\biggl(R_{\mu\nu}^{ab}-\frac{2\Lambda}{3}e_\mu^ae_\nu^b\biggr)=0.
\end{equation}
The components $(\sigma,E,F)=(\sigma,e,4)$ and $(\sigma,E,F)=(\sigma,e,f)$ of Eq.\ \eqref{eq:EoM-A-W-SSB} yield respectively
\begin{equation}
	G^\sigma_e=0,\qquad\tilde{T}^\sigma_{ef}=0.
\end{equation}
Likewise, the components $(\sigma,E,F)=(\sigma,e,4)$ and $(\sigma,E,F)=(\sigma,e,f)$ of Eq.\ \eqref{eq:EoM-A-MM-SSB} yield respectively
\begin{equation}\label{eq:comments}
	G^\sigma_e=0,\qquad\bar{T}^\sigma_{ef}=0.
\end{equation}

The equations of motion for the two pre-geometric theories under exam, that is Eqs.\ \eqref{eq:EoM-A-W} and \eqref{eq:EoM-phi-W} for the Wilczek theory and Eqs.\ \eqref{eq:EoM-A-MM} and \eqref{eq:EoM-phi-MM} for the MacDowell--Mansouri theory, are a set of forty-five coupled second-order nonlinear partial differential equations for the forty independent components of $A$ and the five independent components of $\phi$ (but first-order and linear in $\phi$ in the case of the MacDowell--Mansouri theory). It is important to observe that, just as the respective Lagrangian densities are invariant under $SO(1,4)$ or $SO(2,3)$ before the SSB and $SO(1,3)$ after that, so are the equations of motion of both theories. At the level of equations of motion, the breaking of the original gauge symmetry is signalled by the appearance of the $4$\textsuperscript{th} internal space direction in both equations \eqref{eq:EoM-A-W-SSB} and \eqref{eq:EoM-A-MM-SSB}, meaning that they are covariant only under the residual gauge symmetry of the Lorentz group.

The equations of motion for pre-geometric theories describe the dynamics of three degrees of freedom (see Ref.\ \cite{addazi:hamiltonian} for the details), while the Einstein--Cartan theory contains only two degrees of freedom. In recovering the latter from the former as shown in this section, the third degree of freedom is not lost -- for that would be an inconsistency. Rather, it is only hidden from view through the choice to write, in Eq.\ \eqref{eq:SSB}, the Higgs-like field as equal to its vacuum expectation value in the spontaneously broken phase. This amounts to considering its quantum fluctuations around $v$ as negligible after the SSB, given that such scalar field is expected to be supermassive \cite{addazi:pre-geometry}. Therefore, pre-geometric theories of gravity are different from the Einstein--Cartan theory at a dynamical level precisely because of the dynamics of a new scalar field, which is expected to be frozen out near the Planck scale; only at lower energies, the Einstein--Cartan theory is then recovered from the pre-geometric framework as an effective and emergent theory of gravity. When the quantum fluctuations of $\phi$ are relevant, instead, the pre-geometric theories recover a scalar-tensor metric theory of gravity after the SSB, which is a modified version of the Einstein--Cartan theory with an additional degree of freedom. The Planck scale thus represents the energy regime at which pre-geometric effects become observable in gravitational phenomena.

\section{Emergent Cosmology in vacuum}\label{sec:4}
Solutions of the Einstein field equations are notoriously hard to be found. Solutions of the equations of motion for pre-geometric theories are not any easier to come by, because of their high nonlinearity. For this reason, in this section we will look for what is arguably the simplest possible solution: one describing a pre-geometric universe which is spatially homogenous and isotropic as well as empty.\footnote{In the context of pre-geometric theories of gravity, an empty spacetime is one in which only the pre-geometric fields $A$ and $\phi$ are present. This means that, in addition to $S_\textup{W}$ or $S_\textup{MM}$, the total action will include also kinetic or potential terms for $A$ and $\phi$. Possible examples for constructing these terms in the unbroken phase were shown in Ref.\ \cite{addazi:pre-geometry}. Given that both their form and their relevance is not known in a definitive way yet (see Refs.\ \cite{wilczek:gauge,addazi:pre-geometry,addazi:hamiltonian,addazi:topological} for discussions on different approaches), in this section we omit all such terms in the search for solutions of the equations of motion in vacuum.} For definiteness, we will select the de Sitter group $SO(1,4)$ as the gauge group of each theory, so the internal space metric signature will be $(-,+,+,+,+)$. The implementation of symmetries in the search for solutions of pre-geometric theories can be achieved just as it is done for geometric theories: after constraining the dependence on coordinates of the nonzero components of all (pre-)geometric fields, these are plugged into the equations of motion, whose solution then determines the functional form of such components. Unfortunately, when this scheme is applied to pre-geometric theories, the absence of a metric structure makes it harder to formulate symmetry arguments for determining a priori the nonzero components of the pre-geometric fields. Nonetheless, the correspondence principle can provide some pivotal assistance in this regard, at least as a form of a posteriori justification. In the present context, the correspondence principle can be understood as the requirement that any tensor defined in the unbroken phase of a pre-geometric theory reduces to the analogous quantity of a classical theory of gravity in the spontaneously broken phase, and so in the low-energy limit \cite{addazi:pre-geometry}.

In pre-geometric theories, spatial homogeneity can be implemented by demanding that all pre-geometric fields depend (at most) on the time coordinate only, while spatial isotropy can be implemented by requiring that all the (external or internal) spatial components of any pre-geometric field are equal. The ansatz we make for the gauge field is that only the following components are nonzero, for a total of three independent functions of time:
\begin{equation}\label{eq:ansatz}
	A_0^{04}\equiv E_t(t),\qquad A_i^{i4}\equiv E_s(t),\qquad A_i^{0i}\equiv\Omega(t).
\end{equation}
As for the Higgs-like field, we work in the unitary gauge (the same choice made for the analysis of the gravitational Higgs mechanism in Ref.\ \cite{addazi:pre-geometry}) and write $\phi^A\equiv\Phi(t)\delta_4^A$. Once all of the above symmetries are taken into account, the number of independent equations of motion for both theories reduces to four: the components $(\sigma,E,F)=(0,0,4)$, $(\sigma,E,F)=(i,i,4)$ and $(\sigma,E,F)=(i,0,i)$ of the equations of motion for $A$ and the component $(E)=(4)$ of the equations of motion for $\phi$. Therefore, in each theory there are four independent equations of motion and four independent unknown functions.

For the Wilczek theory, the three independent equations stemming from Eq.\ \eqref{eq:EoM-A-W} are respectively
\begin{subequations}
	\begin{gather}
		\Omega^2-2E_s^2=0,\\
		\Omega^2E_t-2E_s(3E_tE_s-\dot{\Omega})=0,\\
		3E_s\dot{\Phi}-2(\Omega E_t-\dot{E}_s)\Phi=0.
	\end{gather}
\end{subequations}
These can be solved in terms of one of the unknown functions, say $E_s$, to yield\footnote{Nota bene: the double signs found in this section, in which the gauge group is chosen to be $SO(1,4)$, come from taking a square root and should not be confused with those found in the other sections, which instead refer to the two possible choices for the gauge group of the pre-geometric theories as declared in Sec.\ \ref{sec:1}.}
\begin{equation}\label{eq:sol-W}
	\Omega=\pm\sqrt{2}\abs{E_s},\qquad E_t=\pm\frac{\sqrt{2}}{2}\frac{\dot{E}_s}{\abs{E_s}},\qquad\Phi=V,
\end{equation}
where $V$ is a constant. Since $\Phi$ is constant on-shell, that is in the space of solutions respecting the symmetries of the problem, then $\phi^4$ is no longer a local degree of freedom of the theory and the corresponding Euler--Lagrange equation obtained from Eq.\ \eqref{eq:EoM-phi-W} is not useful. As a result, the function $E_s$ remains undetermined.

In the case of the MacDowell--Mansouri theory, the three independent equations coming from Eq.\ \eqref{eq:EoM-A-MM} are respectively
\begin{subequations}
	\begin{gather}
		\Omega^2-E_s^2=0,\\
		\Omega^2E_t-E_s(3E_tE_s-2\dot{\Omega})=0,\\
		(\Omega^2-E_s^2)\dot{\Phi}+2E_s(\Omega E_t-\dot{E}_s)\Phi=0.\label{eq:EoM-MM-3}
	\end{gather}
\end{subequations}
This time, solving in terms of $E_s$ yields
\begin{equation}\label{eq:sol-MM}
	\Omega=\pm\abs{E_s},\qquad E_t=\pm\frac{\dot{E}_s}{\abs{E_s}},
\end{equation}
leaving $\Phi$ undetermined as the Eq.\ \eqref{eq:EoM-MM-3} is automatically satisfied by the solutions \eqref{eq:sol-MM}. These solutions automatically satisfy the component $(E)=(4)$ of the equations of motion \eqref{eq:EoM-phi-MM} too, implying that also $E_s$ remains undetermined.

The choice to motivate the ansatz \eqref{eq:ansatz} a posteriori, rather than a priori, aims at stressing the fact that any solution of the equations of motion for a pre-geometric theory is in principle unrelated to gravity -- it is just a solution of a gauge theory of fields under certain symmetry assumptions. That said, we now proceed to show that the SSB of the solutions \eqref{eq:sol-W} and \eqref{eq:sol-MM} for their respective pre-geometric theories does reproduce an exact solution of the Einstein--Cartan theory -- which is precisely the one respecting the geometric equivalent of the symmetries imposed in the pre-geometric framework.

In the absence of matter, and thus of spin currents, the Cartan field equations \eqref{eq:cartan-field-eq} imply that the torsion of spacetime is absent too. As a consequence, the affine connection $\Gamma$ coincides with the Levi-Civita connection $\mathring{\Gamma}[e,\partial e]$, which allows to compute the corresponding spin connection as
\begin{equation}\label{eq:spin-levi-civita}
	\omega_{b\mu}^a=e_\nu^a\partial_\mu e_b^\nu+e_\nu^a\mathring{\Gamma}_{\rho\mu}^\nu e_b^\rho.
\end{equation}
The most recent cosmological measurements \cite{planck:planck} show that the late Universe is spatially homogenous and isotropic on large enough scales; furthermore, its spatial geometry is flat and its accelerated expansion is compatible with the effect of a positive cosmological constant ($\Lambda>0$). In the Einstein--Cartan theory, such spacetime is described by a flat FLRW metric, which in Cartesian coordinates reads
\begin{equation}\label{FLRW}
	g_{\mu\nu}=\diag(-1,a^2(t),a^2(t),a^2(t)).
\end{equation}
Via the formulae \eqref{eq:soldered} and \eqref{eq:spin-levi-civita}, this translates into
\begin{equation}\label{eq:geometry}
	\begin{split}
		e_0^0=1,\qquad e_i^i=a(t),\qquad\omega_i^{0i}=\dot{a}(t),
	\end{split}
\end{equation}
with all other components of $e$ or $\omega$ vanishing identically. In light of the identifications \eqref{eq:SSB}, we can thus see the correspondence between the pre-geometric ansatz \eqref{eq:ansatz} and the equivalent symmetries assumed in the gravitational theory (Eq.\ \eqref{eq:geometry}), as
\begin{equation}\label{eq:SSB-cosmology}
	E_t\xrightarrow{SSB}me_0^0,\qquad E_s\xrightarrow{SSB}me_i^i,\qquad\Omega\xrightarrow{SSB}\omega_i^{0i}.
\end{equation}
The solution of the vacuum Einstein field equations with a positive cosmological constant is the de Sitter space \cite{mukhanov:cosmology}, whose `cosmological' patch can be expressed as a flat FLRW spacetime \eqref{FLRW} with a scale factor given by
\begin{equation}\label{eq:a}
	a(t)=\exp(\sqrt{\frac{\Lambda}{3}}t),
\end{equation}
which in turn implies that
\begin{equation}
	\begin{split}\label{eq:geometric-sol}
		e_0^0=1,\qquad e_i^i=\exp(\sqrt{\frac{\Lambda}{3}}t),\qquad\omega_i^{0i}=\sqrt{\frac{\Lambda}{3}}\exp(\sqrt{\frac{\Lambda}{3}}t).
	\end{split}
\end{equation}
Going back to the solutions \eqref{eq:sol-W} and \eqref{eq:sol-MM}, after the SSB their form becomes respectively
\begin{equation}
\omega_i^{0i}=\sqrt{2}\abs{me_i^i},\qquad me_0^0=\frac{\sqrt{2}}{2}\frac{m\dot{e}_i^i}{\abs{me_i^i}}
\end{equation}
and
\begin{equation}
\omega_i^{0i}=\abs{me_i^i},\qquad me_0^0=\frac{m\dot{e}_i^i}{\abs{me_i^i}},
\end{equation}
where positive signs were chosen without loss of generality. By substituting the respective expressions for the emergent cosmological constant displayed in Eqs.\ \eqref{eq:emergent-constants-W} and \eqref{eq:emergent-constants-MM}, both theories can finally be seen to yield the same result in the spontaneously broken phase:
\begin{equation}\label{eq:pre-geometric-sol}
	\omega_i^{0i}=\sqrt{\frac{\Lambda}{3}}e_i^i,\qquad e_0^0=\sqrt{\frac{3}{\Lambda}}\frac{\dot{e}_i^i}{\abs{e_i^i}}.
\end{equation}
We can now compare the pre-geometric results \eqref{eq:pre-geometric-sol} in the spontaneously broken phase with the geometric ones \eqref{eq:geometric-sol}: this shows that, for both pre-geometric theories with gauge group $SO(1,4)$, the solution for an empty universe that is spatially homogenous and isotropic correctly predicts the relations between the geometric quantities of the corresponding solution in the gravitational theory, i.e.\ the de Sitter spacetime.\footnote{In the analysis presented in this section, for simplicity it was tacitly assumed that the time coordinate $t$ is the same in the pre-geometric and metric theories; but this need not be the case, as general covariance holds true in both frameworks. To check that this is correct, one can carry out the derivation of the pre-geometric solution with a different time coordinate, say $\tilde{t}$. Then, the chain rule implies that the solution \eqref{eq:pre-geometric-sol} gets modified as
\begin{equation}\label{eq:t-tilde-pre-geometry}
	e_0^0=\sqrt{\frac{3}{\Lambda}}\frac{\partial_{\tilde{t}}e_i^i}{\abs{e_i^i}}=\sqrt{\frac{3}{\Lambda}}\frac{\dot{e}_i^i}{\abs{e_i^i}}\pdv{t(\tilde{t})}{\tilde{t}}.
\end{equation}
Moreover, a change of variables $t\rightarrow\tilde{t}$ transforms the metric for the de Sitter spacetime, and thus its tetrad fields in Eq.\ \eqref{eq:geometric-sol}, as follows:
\begin{equation}\label{eq:t-tilde-geometry}
	e_0^0=1,\quad e_i^i=\exp(\sqrt{\frac{\Lambda}{3}}t)\rightarrow e_0^0=\pdv{t(\tilde{t})}{\tilde{t}},\quad e_i^i=\exp[\sqrt{\frac{\Lambda}{3}}t(\tilde{t})].
\end{equation}
Comparing Eqs.\ \eqref{eq:t-tilde-pre-geometry} and \eqref{eq:t-tilde-geometry} proves that the pre-geometric theories correctly reproduce the relation between the tetrads $e_0^0$ and $e_i^i$ after the SSB also with the new time coordinate $\tilde{t}$.} Both pre-geometric solutions do not specify the form of one of the three independent unknown functions that assume a geometric meaning after the SSB; in the present analysis, this is the function $E_s$, which determines the scale factor $a$ in the spontaneously broken phase. This is consistent with time-reparametrisation invariance, which in metric theories allows to fix the function $e_0^0$ by freely redefining the time coordinate (due to the invariance of the line element). In pre-geometric theories, on the other hand, the redefinition of the time coordinate does not reflect on a fixing of the field $A_0^{04}$ because of the absence of a metric structure. In any case, if the gauge-fixing condition $E_t(t)=m$, which gives $e_0^0\xrightarrow{SSB}1$, is applied to the solution \eqref{eq:pre-geometric-sol}, then one can solve for $e_i^i\xleftarrow{SSB}E_s/m$ and the exponential form of Eq.\ \eqref{eq:a} for the scale factor is correctly recovered in both pre-geometric theories.

The only qualitative difference between the solutions \eqref{eq:sol-W} for the Wilczek theory and \eqref{eq:sol-MM} for the MacDowell--Mansouri theory is in the behaviour of the Higgs-like field. By identifying the constant value of $\phi^4$ in Eq.\ \eqref{eq:sol-W} with its vacuum expectation value, that is $V\equiv v$, one can see that the unbroken phase is not allowed in the Wilczek theory under the symmetry conditions taken into consideration; in other words, the emergence of a metric structure and the recovery of the Einstein--Cartan theory are inevitable for an empty universe which is spatially homogenous and isotropic. In the case of the MacDowell--Mansouri theory, on the other hand, the evolution of the Higgs-like field is unconstrained by the symmetries of the problem; therefore, the SSB of the theory remains subject to the dynamics of $\phi$ and the de Sitter universe admits a pre-geometric phase not only in a formal but also in a physical sense. In any case, it must be kept in mind that the symmetries imposed in the search for these solutions correspond to the idealised scenario of the complete absence of inhomogeneities. Because of that, the pre-geometric solution for the de Sitter spacetime will be relevant in the Wilczek theory too as soon as any small perturbations are taken into account.

Since the de Sitter spacetime is a good description, in the first approximation, also for the inflationary epoch of the Universe \cite{mukhanov:cosmology}, its pre-geometric version provides a possible solution to the problem of the Big Bang singularity. According to the results \eqref{eq:SSB-cosmology} and \eqref{eq:geometric-sol}, in fact, in both pre-geometric theories no singularities arise both before and after the SSB: in the spontaneously broken phase, the exponential form of the geometric quantities is a smooth function for any value of the time coordinate; then in the unbroken phase, the pre-geometric fields remain likewise non-singular at any time, including the value that would correspond to the Big Bang in the geometric phase of spacetime. At a critical energy near the Planck scale \cite{addazi:pre-geometry}, and thus at a critical time in the past evolution of the Universe, any (possibly singular) metric structure of spacetime is thus superseded by the dynamics of a non-singular theory à la Yang--Mills, as the fundamental gauge symmetry of spacetime is restored in the ultra-high-energy limit. Pre-geometric theories provide a simple resolution of the issue posed by the BGV theorem too. According to this kinematic argument \cite{borde:incomplete}, any universe that, at least on average, has always been expanding must also be geodesically incomplete in the past, independently of the metric theory of gravity governing its evolution. This theorem only applies to a classical spacetime, i.e.\ one endowed with a metric structure, and does not imply the existence of the Big Bang singularity, rather only that a metric expansion cannot be infinite in the past. Therefore, the spacetime boundary of the early Universe simply serves as the indication of when the classical approximation breaks down \cite{carroll:something}; in pre-geometric theories, this is signalled by the phase transition to a pre-geometric universe in the ultra-high-energy regime, as the geometric picture ceases to be valid. In this scenario, the Big Bang should then be thought of not as the (impossible) beginning of the Universe, but rather as the time when its geometric expansion started after a phase transition from a pre-geometric spacetime.

\section{Pre-geometric dynamics with matter}\label{sec:5}
We now turn to the problem of the matter coupling in pre-geometric theories of gravity. The Lagrangian treatment was previously sketched out in Ref.\ \cite{addazi:pre-geometry}, elucidating a possible pre-geometric generalisation for the total Lagrangian density $\mathcal{L}_m$.\footnote{Despite adopting the same notation for both the unbroken and the spontaneously broken phases, whether in this section $\mathcal{L}_m$ refers to its geometric or pre-geometric formulation will be clear from the context.} In this work we develop a more general approach based on the variational principle as well as the correspondence principle (which was stated at the beginning of Sec.\ \ref{sec:4}). We posit that there exists a coupling between matter and the pre-geometry of spacetime and that, following the correspondence principle, this coupling to pre-geometry correctly reproduces that to geometry after the SSB.

The total action for the pre-geometric fields $A$ and $\phi$ in the Wilczek theory and $n$ matter fields $\psi_n$ is
\begin{equation}\label{eq:S-pre-geometry-W}
	S=S_\textup{W}[\phi,\partial\phi,A,\partial A]+S_m[\phi,\partial\phi,A,\psi_n,\partial\psi_n],
\end{equation}
with $S_\textup{W}$ defined in Eqs.\ \eqref{eq:SSB-W} and \eqref{eq:L-W}. The functional dependence stated for $S_m$ in the unbroken phase is consistent with the observation made in Sec.\ \ref{sec:2} that, in the spontaneously broken phase, all terms of $S$ containing $\partial\omega\xleftarrow{SSB}\partial A$ belong to the gravitational part $S_\textup{EC}\xleftarrow{SSB}S_\textup{W}$ of the action and not to the matter part $S_m$. The matter action is defined as
\begin{equation}\label{eq:S-matter}
	S_m=\int\frac{\abs{J}}{24v^5m^4}\mathcal{L}_m\,d^4x,
\end{equation}
where the normalisation factor\footnote{The constants $v$ and $m$ characterise the spontaneously broken phase of pre-geometric theories. In the treatment of this section we choose to display such constants in $S_m$ already in the unbroken phase for the sake of clarity, even though they can also be absorbed in a redefinition of the coupling constants for the matter fields $\psi_n$ present in $\mathcal{L}_m$.} is introduced for dimensional reasons and chosen according to the correspondence principle, as
\begin{equation}
	\begin{split}
		J&\equiv\epsilon_{ABCDE}\epsilon^{\mu\nu\rho\sigma}\nabla_\mu\phi^A\nabla_\nu\phi^B\nabla_\rho\phi^C\nabla_\sigma\phi^D\phi^E\\
		&\xrightarrow{SSB}-24v^5m^4e.
	\end{split}
\end{equation}

The field equations obtained from applying the variational principle with respect to $A$ to the action \eqref{eq:S-pre-geometry-W} can be referred to as the pre-geometric Einstein--Cartan field equations:
\begin{equation}\label{eq:EC-eqs}
	\mathcal{G}_{AB}^\mu=\mathcal{M}_\textup{P}^{-2}\mathcal{T}_{AB}^\mu,
\end{equation}
where the pre-geometric Einstein--Cartan tensor is defined as
\begin{equation}
	\begin{split}
		\mathcal{G}_{AB}^\mu\equiv\frac{12v^2m^3}{\abs{J}}\epsilon^{\mu\nu\rho\sigma}[&\epsilon_{CDEF[A}\eta_{B]G}(F_{\nu\rho}^{CD}\phi^G\\
		&-2A_\nu^{CG}\nabla_\rho\phi^D)\nabla_\sigma\phi^E\phi^F\\
		&+\epsilon_{ABCDE}\partial_\sigma(\nabla_\nu\phi^C\nabla_\rho\phi^D\phi^E)],
	\end{split}
\end{equation}
the canonical source tensor is defined as
\begin{equation}
	\mathcal{T}_{AB}^\mu\equiv\frac{2m}{\abs{J}}\fdv{(\abs{J}\mathcal{L}_m)}{A_\mu^{AB}}=\frac{2m}{\abs{J}}\pdv{(\abs{J}\mathcal{L}_m)}{A_\mu^{AB}}
\end{equation}
and the effective coupling constant appearing on the r.h.s.\ is defined as
\begin{equation}\label{eq:SSB-MP-W}
	\mathcal{M}_\textup{P}\equiv\sqrt{-8k_\textup{W}v^3m^2}.
\end{equation}

The reasons for the proposed names of the tensors $\mathcal{G}$ and $\mathcal{T}$ can easily be understood from the analysis of the SSB of the theory. First of all, Eq.\ \eqref{eq:emergent-constants-W} shows that the effective coupling constant yields exactly the reduced Planck mass in the spontaneously broken phase, i.e.\ $\mathcal{M}_\textup{P}\xrightarrow{SSB}M_\textup{P}$. As for the pre-geometric Einstein--Cartan tensor, its components $(\mu,A,B)=(\mu,a,4)$ and $(\mu,A,B)=(\mu,a,b)$ are respectively proportional to the Einstein tensor\footnote{Extra care must be taken when computing the SSB of the pre-geometric Einstein--Cartan tensor if one of its internal indices coincides with the direction singled out by the SSB. The part of $\mathcal{G}$ containing an explicit antisymmetrisation of the internal indices comes from computing $\partial\mathcal{L}_\textup{W}/\partial A_\mu^{AB}$ by making use of the formula (neglecting the dependence on the spacetime coordinates)
\begin{equation}
	\pdv{A_\nu^{CD}}{A_\mu^{AB}}=\delta_\nu^\mu\delta_{[A}^C\delta_{B]}^D,
\end{equation}
where the antisymmetrisation of the internal indices on the r.h.s.\ is due to the gauge field $A$ being an antisymmetric tensor in such indices \cite{itskov:derivative}. In particular, this implies that
\begin{equation}\label{eq:diff-antisymmetric}
	\pdv{A_\nu^{c4}}{A_\mu^{a4}}=\delta_\nu^\mu\delta_{[a}^c\delta_{4]}^4=\frac{1}{2}\delta_\nu^\mu\delta_a^c.
\end{equation}
On the other hand, it is also true that, by means of the identifications of Eq.\ \eqref{eq:SSB}, the mechanism of SSB yields
\begin{equation}
	\pdv{A_\nu^{c4}}{A_\mu^{a4}}\xrightarrow{SSB}\pdv{e_\nu^c}{e_\mu^a}=\delta_\nu^\mu\delta_a^c,
\end{equation}
where in this case the absence of the factor $1/2$ on the r.h.s.\ compared to the result of Eq.\ \eqref{eq:diff-antisymmetric} is due to the tetrad fields $e$ having only one internal index. In other words, in identifying certain components of the gauge field as tetrads after the SSB, the antisymmetric character of the internal indices of $A$ goes missing and this affects the way its differentiation is computed. To obviate this drawback without altering the form of the pre-geometric Einstein--Cartan tensor, we employ the prescription of dropping the explicit antisymmetrisation of the internal indices for the term that contains it (or, equivalently, to multiply it by a factor of $2$) whenever the SSB of $\mathcal{G}$ is computed along the $4$\textsuperscript{th} internal space direction.} and the modified torsion tensor after the SSB:
\begin{equation}
	\qquad\qquad\quad\mathcal{G}_{a4}^\mu\xrightarrow{SSB}2G_a^\mu,\qquad\mathcal{G}_{ab}^\mu\xrightarrow{SSB}m\tilde{T}_{ab}^{\mu}.
\end{equation}
The corresponding components of the canonical source tensor are respectively proportional to the canonical energy-momentum tensor and the canonical spin tensor after the SSB:
\begin{equation}\label{eq:SSB-T}
	\mathcal{T}_{a4}^\mu\xrightarrow{SSB}2\tau_a^\mu,\qquad\mathcal{T}_{ab}^\mu\xrightarrow{SSB}m\sigma_{ab}^{\mu}.
\end{equation}
Putting everything together, the SSB of the pre-geometric Einstein--Cartan field equations \eqref{eq:EC-eqs} yields the Einstein field equations for the components $(\mu,A,B)=(\mu,a,4)$ and the Cartan field equations for the components $(\mu,A,B)=(\mu,a,b)$:
\begin{subequations}
	\begin{gather}
		\mathcal{G}_{a4}^\mu=\mathcal{M}_\textup{P}^{-2}\mathcal{T}_{a4}^\mu\xrightarrow{SSB}G_a^\mu=M_\textup{P}^{-2}\tau_a^\mu,\\
		\mathcal{G}_{ab}^\mu=\mathcal{M}_\textup{P}^{-2}\mathcal{T}_{ab}^\mu\xrightarrow{SSB}\tilde{T}_{ab}^\mu=M_\textup{P}^{-2}\sigma_{ab}^\mu.
	\end{gather}
\end{subequations}
This result can also be visualised more explicitly as\footnote{The form of the pre-geometric Einstein--Cartan field equations with mixed types of indices, i.e.\ both spacetime and tangent space indices, is the most suitable one for capturing the mechanism of SSB. In the unbroken phase, in fact, there exist no tetrads to convert between the two types of indices. Then, in the spontaneously broken phase the $4$\textsuperscript{th} internal space direction is no longer part of the residual gauge symmetry, manifesting an asymmetry with respect to the other internal space directions and thus between the Einstein and the Cartan field equations; this is highlighted also by the fact that it takes multiplication by one or two tetrads respectively to recast those equations in the form with spacetime indices only.}
\begin{widetext}
	\begin{equation}
		\mathcal{G}_{AB}^\mu=\mathcal{M}_\textup{P}^{-2}\mathcal{T}_{AB}^\mu\xrightarrow{SSB}\mqty(\mqty{0&\tilde{T}_{01}^\mu&\tilde{T}_{02}^\mu&\tilde{T}_{03}^\mu\\-\tilde{T}_{01}^\mu&0&\tilde{T}_{12}^\mu&\tilde{T}_{13}^\mu\\-\tilde{T}_{02}^\mu&-\tilde{T}_{12}^\mu&0&\tilde{T}_{23}^\mu\\-\tilde{T}_{03}^\mu&-\tilde{T}_{13}^\mu&-\tilde{T}_{23}^\mu&0}&\rvline&\mqty{G_0^\mu\\G_1^\mu\\G_2^\mu\\G_3^\mu}\\\hline\mqty{-G_0^\mu&-G_1^\mu&-G_2^\mu&-G_3^\mu}&\rvline&0)=M_\textup{P}^{-2}\mqty(\mqty{0&\sigma_{01}^\mu&\sigma_{02}^\mu&\sigma_{03}^\mu\\-\sigma_{01}^\mu&0&\sigma_{12}^\mu&\sigma_{13}^\mu\\-\sigma_{02}^\mu&-\sigma_{12}^\mu&0&\sigma_{23}^\mu\\-\sigma_{03}^\mu&-\sigma_{13}^\mu&-\sigma_{23}^\mu&0}&\rvline&\mqty{\tau_0^\mu\\\tau_1^\mu\\\tau_2^\mu\\\tau_3^\mu}\\\hline\mqty{-\tau_0^\mu&-\tau_1^\mu&-\tau_2^\mu&-\tau_3^\mu}&\rvline&0).
	\end{equation}
\end{widetext}

The analysis for the matter coupling in the MacDowell--Mansouri theory is analogous. The total action in this case is
\begin{equation}\label{eq:S-pre-geometry-MM}
	S=S_\textup{MM}[\phi,A,\partial A]+S_m[\phi,\partial\phi,A,\psi_n,\partial\psi_n],
\end{equation}
with $S_\textup{MM}$ defined in Eqs.\ \eqref{eq:SSB-MM} and \eqref{eq:L-MM} and $S_m$ defined in Eq.\ \eqref{eq:S-matter}. The field equations obtained from applying the variational principle with respect to $A$ to the action \eqref{eq:S-pre-geometry-MM} are again the pre-geometric Einstein--Cartan field equations \eqref{eq:EC-eqs}, but with a different definition for $\mathcal{G}$ and $\mathcal{M}_\textup{P}$. This time, the pre-geometric Einstein--Cartan tensor is found to be
\begin{equation}
	\begin{split}
		\mathcal{G}_{AB}^\mu\equiv\pm\frac{6v^4m^3}{\abs{J}}\epsilon^{\mu\nu\rho\sigma}[&2\epsilon_{CDEF[A}A_{B]\nu}^CF_{\rho\sigma}^{DE}\phi^F\\
		&-\epsilon_{ABCDE}\partial_\sigma(F_{\nu\rho}^{CD}\phi^E)]
	\end{split}
\end{equation}
and the effective coupling constant is
\begin{equation}\label{eq:SSB-MP-MM}
	\mathcal{M}_\textup{P}\equiv\sqrt{\pm32k_\textup{MM}vm^2}.
\end{equation}

As for the spontaneously broken phase, again we find that $\mathcal{M}_\textup{P}\xrightarrow{SSB}M_\textup{P}$. The SSB of $\mathcal{G}$ is given instead by
\begin{equation}
	\mathcal{G}_{a4}^\mu\xrightarrow{SSB}2G_a^\mu,\qquad\mathcal{G}_{ab}^\mu\xrightarrow{SSB}m\bar{T}_{ab}^{\mu}.
\end{equation}
Given that the canonical source tensor is the same as that of the Wilczek theory in both the unbroken and the spontaneously broken phases, the SSB of the pre-geometric Einstein--Cartan field equations in this case yields
\begin{subequations}
	\begin{gather}
		\mathcal{G}_{a4}^\mu=\mathcal{M}_\textup{P}^{-2}\mathcal{T}_{a4}^\mu\xrightarrow{SSB}G_a^\mu=M_\textup{P}^{-2}\tau_a^\mu,\\
		\mathcal{G}_{ab}^\mu=\mathcal{M}_\textup{P}^{-2}\mathcal{T}_{ab}^\mu\xrightarrow{SSB}\bar{T}_{ab}^\mu=M_\textup{P}^{-2}\sigma_{ab}^\mu;
	\end{gather}
\end{subequations}
that is,
\begin{widetext}
	\begin{equation}
		\mathcal{G}_{AB}^\mu=\mathcal{M}_\textup{P}^{-2}\mathcal{T}_{AB}^\mu\xrightarrow{SSB}\mqty(\mqty{0&\bar{T}_{01}^\mu&\bar{T}_{02}^\mu&\bar{T}_{03}^\mu\\-\bar{T}_{01}^\mu&0&\bar{T}_{12}^\mu&\bar{T}_{13}^\mu\\-\bar{T}_{02}^\mu&-\bar{T}_{12}^\mu&0&\bar{T}_{23}^\mu\\-\bar{T}_{03}^\mu&-\bar{T}_{13}^\mu&-\bar{T}_{23}^\mu&0}&\rvline&\mqty{G_0^\mu\\G_1^\mu\\G_2^\mu\\G_3^\mu}\\\hline\mqty{-G_0^\mu&-G_1^\mu&-G_2^\mu&-G_3^\mu}&\rvline&0)=M_\textup{P}^{-2}\mqty(\mqty{0&\sigma_{01}^\mu&\sigma_{02}^\mu&\sigma_{03}^\mu\\-\sigma_{01}^\mu&0&\sigma_{12}^\mu&\sigma_{13}^\mu\\-\sigma_{02}^\mu&-\sigma_{12}^\mu&0&\sigma_{23}^\mu\\-\sigma_{03}^\mu&-\sigma_{13}^\mu&-\sigma_{23}^\mu&0}&\rvline&\mqty{\tau_0^\mu\\\tau_1^\mu\\\tau_2^\mu\\\tau_3^\mu}\\\hline\mqty{-\tau_0^\mu&-\tau_1^\mu&-\tau_2^\mu&-\tau_3^\mu}&\rvline&0).
	\end{equation}
\end{widetext}

We note in passing that it is possible to define also the quantity
\begin{equation}
	\mathcal{T}_A\equiv\frac{1}{\abs{J}}\fdv{(\abs{J}\mathcal{L}_m)}{\phi^A},
\end{equation}
though its meaning is not readily understood via the correspondence principle.

\section{Emergent Cosmology with matter}\label{sec:6}
Having laid out the formalism of the matter coupling in the pre-geometric theories, in this section we revisit the application to a pre-geometric universe that is spatially homogenous and isotropic (as studied in Sec.\ \ref{sec:4}), but add matter to it. Dropping the assumption of emptiness comes with the price of reduced spacetime symmetries and higher complexity of the field equations, especially in the unbroken phase.

We employ the same symmetry conditions and notation for the pre-geometric fields as done in Sec.\ \ref{sec:4} (see Eq.\ \eqref{eq:ansatz}), as well as the same gauge group $SO(1,4)$. The components $(\mu,A,B)=(0,0,4)$, $(\mu,A,B)=(i,i,4)$ and $(\mu,A,B)=(i,0,i)$ of the pre-geometric Einstein--Cartan field equations \eqref{eq:EC-eqs} for the Wilczek theory are thus found to be respectively
\begin{subequations}
	\begin{gather}
		-3v^2m^3\frac{\Omega^2-2E_s^2}{\sign{(E_s\Phi)}E_s^2\Phi^2\abs{E_t}}=\mathcal{M}_\textup{P}^{-2}\frac{\mathcal{T}_{04}^0}{2},\\
		-v^2m^3\frac{\Omega^2E_t-2E_s(3E_tE_s-\dot{\Omega})}{\sign{(\Phi)}\Phi^2\abs{E_tE_s^3}}=\mathcal{M}_\textup{P}^{-2}\frac{\mathcal{T}_{i4}^i}{2},\\
		v^2m^2\frac{3E_s\dot{\Phi}-2(\Omega E_t-\dot{E}_s)\Phi}{\sign{(E_s)}E_s^2\abs{E_t\Phi^3}}=\mathcal{M}_\textup{P}^{-2}\frac{\mathcal{T}_{0i}^i}{m},
	\end{gather}
\end{subequations}
while for the MacDowell--Mansouri theory they are respectively
\begin{subequations}
	\begin{gather}
		-3v^4m^3\frac{\Omega^2-E_s^2}{\sign{(E_s\Phi)}E_s^2\Phi^4\abs{E_t}}=\mathcal{M}_\textup{P}^{-2}\frac{\mathcal{T}_{04}^0}{2},\\
		-v^4m^3\frac{\Omega^2E_t-E_s(3E_tE_s-2\dot{\Omega})}{\sign{(\Phi)}\Phi^4\abs{E_tE_s^3}}=\mathcal{M}_\textup{P}^{-2}\frac{\mathcal{T}_{i4}^i}{2},\\
		-v^4m^2\frac{(\Omega^2-E_s^2)\dot{\Phi}+2E_s(\Omega E_t-\dot{E}_s)\Phi}{\abs{E_tE_s^3\Phi^5}}=\mathcal{M}_\textup{P}^{-2}\frac{\mathcal{T}_{0i}^i}{m}.
	\end{gather}
\end{subequations}

This time we limit ourselves to verifying that these pre-geometric field equations correctly reproduce the corresponding field equations of the gravitational theory in the spontaneously broken phase. In the Einstein--Cartan theory, the geometry of spacetime is again described by the quantities of Eq.\ \eqref{eq:geometry}, with $\Lambda>0$, i.e.\ a flat FLRW metric with Cartesian coordinates. As for the matter content, we pick the simplest case of a spinless perfect fluid with energy density $\rho$ and isotropic pressure $p$, whose canonical energy-momentum tensor is
\begin{equation}\label{eq:perfect-fluid}
	\tau_a^\mu=e_a^\nu\tau_\nu^\mu=\diag\Bigl(-\rho,\frac{p}{a},\frac{p}{a},\frac{p}{a}\Bigr).
\end{equation}
The results \eqref{eq:SSB}, \eqref{eq:emergent-constants-W}, \eqref{eq:emergent-constants-MM}, \eqref{eq:SSB-MP-W}, \eqref{eq:SSB-T} and \eqref{eq:SSB-MP-MM} allow to compute the SSB of the pre-geometric field equations of this section for both the Wilczek and the MacDowell--Mansouri theories; in particular, the SSB for the components $(\mu,A,B)=(0,0,4)$ and $(\mu,A,B)=(i,i,4)$ is given in both cases respectively by
\begin{subequations}
	\begin{gather}
		\frac{3(\omega_i^{0i})^2-\Lambda(e_i^i)^2}{e_0^0(e_i^i)^2}=-M_\textup{P}^{-2}\tau_0^0,\\
		\frac{e_0^0(\omega_i^{0i})^2-\Lambda e_0^0(e_i^i)^2+2e_i^i\dot{\omega}_i^{0i}}{e_0^0(e_i^i)^3}=-M_\textup{P}^{-2}\tau_i^i.
	\end{gather}
\end{subequations}
Finally, substituting the definitions \eqref{eq:geometry} and \eqref{eq:perfect-fluid} from the Einstein--Cartan theory leads in both pre-geometric theories exactly to the Friedmann equations \cite{mukhanov:cosmology} for a spatially flat universe with a positive cosmological constant in the spontaneously broken phase:
\begin{subequations}
	\begin{gather}
		\biggl(\frac{\dot{a}}{a}\biggr)^2-\frac{\Lambda}{3}=\frac{M_\textup{P}^{-2}}{3}\rho,\\
		2\frac{\ddot{a}}{a}+\biggl(\frac{\dot{a}}{a}\biggr)^2-\Lambda=-M_\textup{P}^{-2}p.
	\end{gather}
\end{subequations}

\section{Conclusions and perspectives}\label{sec:7}
With this work we proved that the Einstein--Cartan theory is correctly recovered from the SSB of pre-geometric theories of gravity at all theoretical levels: not only in terms of Lagrangian and Hamiltonian formulations (as shown in Refs.\ \cite{addazi:pre-geometry} and \cite{addazi:hamiltonian} respectively), but also in terms of equations of motion and their solutions (at least in one case). In these theories à la Yang--Mills, the SSB reduces the fundamental gauge symmetry of spacetime to the one observed at low energies, i.e.\ the Lorentz symmetry; the main by-products of this mechanism are then the emergence of a metric structure for spacetime as well as the emergence of the gravitational interaction in its geometric form. The field equations for pre-geometric theories were analysed both without and with matter. The difference between the pre-geometric and the metric theories is independent of the presence of matter: both without or with matter, the fundamental field equations are different (and with a different number of degrees of freedom); and both without or with matter, the fundamental field equations reduce to those of the Einstein--Cartan theory after the SSB (if the quantum fluctuations of the Higgs-like field are negligible). In the case where matter is present, this reduction via SSB requires understanding how matter couples to the pre-geometry of spacetime, which can be achieved in a general way by means of the correspondence principle. The pre-geometric field equations can be obtained by varying the total action (for pre-geometry and matter) with respect to the gauge field of the theory à la Yang--Mills. The ensuing equations encapsulate both the Einstein and the Cartan field equations for gravitation after the SSB. This result points at a common theoretical origin for the coupling between all quantities describing a geometric spacetime on one side, that is the metric and the affine connection, and matter on the other side, that is the energy-momentum and spin properties. Specifically, in the unbroken phase all features of matter can be ascribed to a single tensor, the canonical source tensor; in the spontaneously broken phase, in fact, its components contain both the canonical energy-momentum tensor and the canonical spin tensor.

Until the SSB eventually occurs (below a critical ultra-high energy), there is no reason to expect the pre-geometric field equations to be even only related to those of the Einstein--Cartan theory: not only are they very different at a mathematical level (see comments after Eq.\ \eqref{eq:comments}), but they also contain more degrees of freedom, which allow for a richer solution space. Moreover, a priori there is no indication that the pre-geometric solutions imply any metric interpretation at all -- that is possible only once the SSB occurs, indeed. This motivates the search for solutions of the fundamental pre-geometric field equations, under suitable symmetry assumptions; only after that, their consistency with a metric theory of gravity can be checked in the spontaneously broken phase. In the unbroken phase, on the other hand, one is in principle oblivious to the form, as well as the meaning, of the pre-geometric solutions for the gauge field of the theory. A first exact solution of the pre-geometric Einstein--Cartan field equations was then presented and discussed. As shown by studying its SSB, this solution can be considered as a pre-geometric de Sitter universe in the unbroken phase; that is, in the spontaneously broken phase it gives rise to an empty spacetime that is spatially homogenous and isotropic, as well as spatially flat and with a positive cosmological constant. Since the SSB is expected to happen near the Planck scale and the de Sitter universe is a good approximation for the inflationary epoch, a direct consequence of the pre-geometric solution is the possibility of explaining away the Big Bang singularity. In the ultra-high-energy limit, in fact, the fundamental gauge symmetry of spacetime is restored and the Einstein--Cartan theory is superseded by a pre-geometric one whose solution is singularity-free, annulling any notions of geometric singularities or geodesic incompleteness. Therefore, in this context the Big Bang singularity can be seen simply as a wrong extrapolation made from the predictions of a low-energy gravitational theory applied beyond its domain of validity. Rather, the Big Bang could be referred to as the time when a metric structure emerges from a pre-geometric universe and the geometric expansion of spacetime begins.

One question that remains open is about the pre-geometric field equations obtained from the variation of the total action with respect to the Higgs-like field. The correspondence principle, in fact, offers no direct interpretation for them. In particular, it will be interesting to explore in future works the role of a possible matter coupling in such equations, given that it is defined in a distinct way from that of the canonical source tensor. As for the most intriguing application of the present study, this arguably comes from understanding the role of the quantum fluctuations of the Higgs-like field around its vacuum expectation value: these survive in the form of a supermassive scalar field after the SSB \cite{addazi:pre-geometry}, giving rise to what is effectively a scalar-tensor metric theory of gravity in the spontaneously broken phase \cite{addazi:hamiltonian}. The dynamical consequences of this additional degree of freedom can spark a search for deviations from Einstein gravity in different areas of physical Cosmology, including Inflation, cosmological perturbations, gravitational waves and dark matter. In addition to that, the pre-geometric framework is consistent with the experimental signatures of torsion that have recently received plenty of attention in the literature, especially with reference to cosmological observations \cite{cai:f(T),hehl:spin,alexander:spectrum,alexander:curvaton,addazi:dark,addazi:ekpyrotic,addazi:emergent,popławski:torsion,basilakos:noether,paliathanasis:solutions,chen:perturbations,capozziello:transition,cai:matter}.\linebreak

{\bf Acknowledgements}.
The author acknowledges the support of Istituto Nazionale di Fisica Nucleare, Sezione di Napoli, Iniziativa Specifica QGSKY. This paper is based upon work from COST Action CA21136 -- Addressing observational tensions in cosmology with systematics and fundamental physics (CosmoVerse), supported by COST (European Cooperation in Science and Technology).

\end{document}